\pdfoutput=1

\documentclass[11pt]{article}

\usepackage[final]{acl}

\usepackage{times}
\usepackage{latexsym}

\usepackage[T1]{fontenc}

\usepackage[utf8]{inputenc}
\usepackage{microtype}

\usepackage{inconsolata}
\usepackage{enumitem}

\usepackage{graphicx}

\usepackage{multirow, graphicx}

\usepackage{tabularray}
\usepackage{xcolor}   
\usepackage{newfloat} 
\usepackage{listings}

\newcommand\myparagraph[1]{
\vskip 0.05in 
\noindent{\bf {#1}}}

\title{ClimRetrieve: A Benchmarking Dataset for Information Retrieval from Corporate Climate Disclosures}

\author{Tobias Schimanski\textsuperscript{\rm 1},
    Jingwei Ni\textsuperscript{\rm 2},
    Roberto Spacey\textsuperscript{\rm 3},
    Nicola Ranger\textsuperscript{\rm 3},
    Markus Leippold\textsuperscript{\rm 1, 4} \\
    \textsuperscript{\rm 1}University of Zurich
    \textsuperscript{\rm 2}ETH Zurich
    \textsuperscript{\rm 3}University of Oxford
    \textsuperscript{\rm 4}Swiss Finance Institute (SFI) \\  
\texttt{tobias.schimanski@df.uzh.ch}
    }

\begin{document}
\maketitle
\begin{abstract}
To handle the vast amounts of qualitative data produced in corporate climate communication, stakeholders increasingly rely on Retrieval Augmented Generation (RAG) systems. However, a significant gap remains in evaluating domain-specific information retrieval -- the basis for answer generation. To address this challenge, this work simulates the typical tasks of a sustainability analyst by examining 30 sustainability reports with 16 detailed climate-related questions. As a result, we obtain a dataset with over 8.5K unique question-source-answer pairs labeled by different levels of relevance. Furthermore, we develop a use case with the dataset to investigate the integration of expert knowledge into information retrieval with embeddings. Although we show that incorporating expert knowledge works, we also outline the critical limitations of embeddings in knowledge-intensive downstream domains like climate change communication.\footnote{All the data and code for this project is available on \url{https://github.com/tobischimanski/ClimRetrieve}.}\footnote{We thank the expert annotators Aysha Emmerson,  Emily Hsu, and Capucine Le Meur for their work on this project.}
\end{abstract}

\section{Introduction} 

\textbf{Motivation.}
Climate change presents the most pressing challenge of our time. The underlying concepts and challenges generate a wealth of information with inherent complexity and interconnectedness. At the same time, most of the data on corporate climate disclosure is qualitative -- hidden in textual statements \citep{Weber2023ClimateCR,sec_rule}. Qualitative disclosures typically include narrative descriptions of climate-related risks, opportunities, strategies, and governance. These are crucial to understanding how a company perceives and manages climate-related issues and their potential impacts on business operations.\footnote{For example, companies must describe the processes they use to identify, assess, and manage these risks and opportunities, as well as the roles of the board and management in these processes.}

Advances in Natural Language Processing (NLP) try to address data structuring and analysis challenges. Specifically, Retrieval-Augmented-Generation (RAG) emerged as a method to address knowledge-intensive questions around climate change \citep{vaghefi2023chatclimate, niCHATREPORTDemocratizingSustainability2023, Colesanti2024CombiningAA}. Despite the growing demand for more precise climate change data \citep{Sietsma2023TheNG}, a significant gap exists in evaluating RAG systems. While researchers have developed methodologies for the automatic evaluation of generated content \citep{chenBenchmarkingLargeLanguage2023, schimanski2024faithful, saadfalcon2024ares}, the preceding crucial phase of information retrieval remains largely unexamined in the context of climate change.\footnote{For a recent solution approach, see \citealp{ni2024dirasefficientllmassistedannotation}.}

\begin{figure}[t]
    \centering	\includegraphics[width=0.45\textwidth]{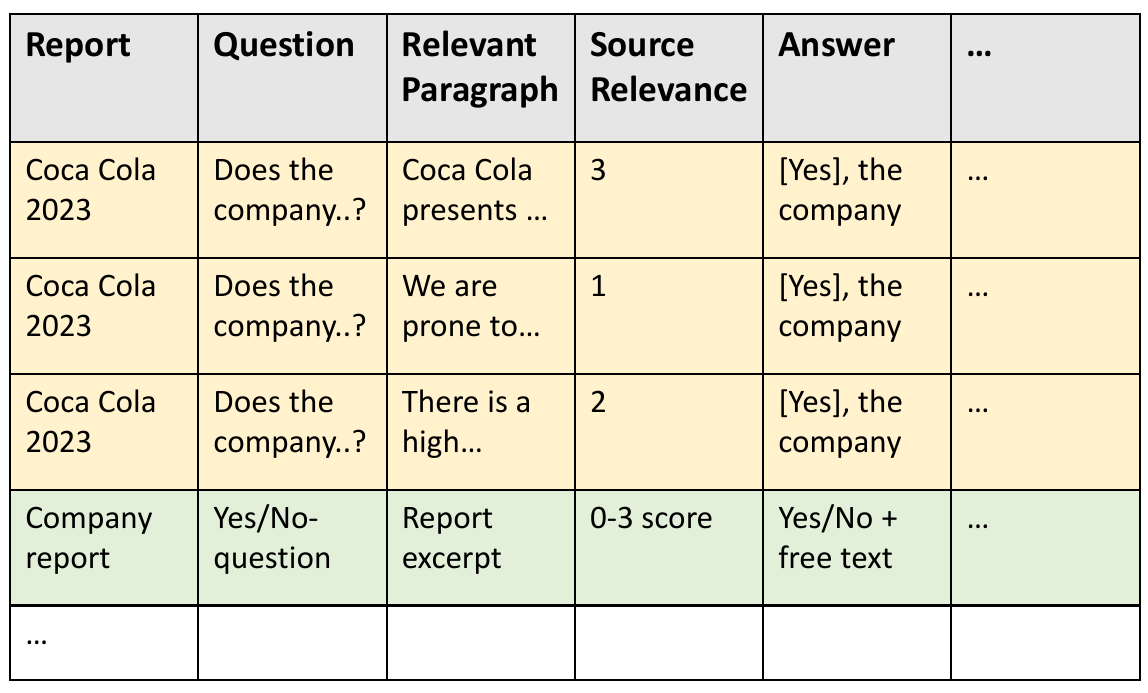}
	\caption{Overview of the core columns of ClimRetrieve.}
	\label{fig:overview_ClimRAG}
 \vspace{-1em}
\end{figure}

\noindent \textbf{Contribution.} 
Therefore, this paper delivers two contributions. First, it introduces a comprehensive expert-annotated dataset for the retrieval and generation part of RAG. The dataset emulates an analyst workflow to answer questions based on the provided documents. Thus, the core data set comprises questions, the corresponding sources recovered from experts, their relevance, and an answer to the question (see Figure \ref{fig:overview_ClimRAG}). Second, we design an experiment to compare human expert annotations with various embedding search strategies. This investigation aims to understand how to integrate expert knowledge into the retrieval process.

\noindent \textbf{Results.} 
We find that SOTA embedding models (on which RAG systems heavily rely) usually fail to effectively reflect domain expertise. This shows that bringing expert knowledge into the retrieval process is a non-trivial task. Thus, we underline the importance of new approaches in information retrieval. This dataset can present a basis for improvement approaches. 

\noindent \textbf{Implications.} 
The implications of our study are significant for both practice and research. Knowledge-intensive downstream domains like climate change are nuanced, and details matter. This paper can significantly help researchers evaluate new RAG systems (e.g., \citealp{niCHATREPORTDemocratizingSustainability2023}) and corporate climate report analysts to obtain useful information for decision-making.

\section{Background} \label{sec:background}
\textbf{Retrieval Augmented Generation (RAG).}
RAG has been widely adopted to mitigate hallucination and enhance application performance \citep{vaghefi2023chatclimate, niCHATREPORTDemocratizingSustainability2023, Colesanti2024CombiningAA}. RAG systems base their answers on external information integrated into the prompt rather than parametric knowledge learned during training \citep{Lewis_2020}. This approach critically shifts the problem from learning the information during training to retrieving the right information and summarizing and arguing over the provided content. Many related projects explore how to evaluate the quality of LLM generation augmented with retrieval \citep{zhang2024raft, saadfalcon2024ares, asai2023selfrag, schimanski2024faithful}. However, how to directly assess the information retrieval thoroughness and precision is still underexplored, especially for specific but important domains like corporate climate disclosure. The only work to date that tries to integrate domain-specific nuances explicitly is \citealp{ni2024dirasefficientllmassistedannotation}.

\noindent \textbf{Climate Change NLP.}
Prior work, specifically before the popularisation of RAG, has mainly worked with BERT-based classifiers to address climate change questions. This ranges from the verification of environmental claims \cite{stammbach2022dataset}, the detection of climate change topics \citep{varini2021climatext}, the verification of facts \citep{diggelmann2021climatefever, leippold2024automated}, the detection of net zero and reduction targets \citep{schimanski-etal-2023-climatebert} or more generally environmental, social and governance texts \citep{SCHIMANSKI2024104979}. Although this provided valuable information on communication patterns, for example, in corporate reporting \citep{BINGLER2024107191, koelbel22ask}, fine-granular, nuanced reasoning analyses were only enabled after the popularization of RAG \citep{niCHATREPORTDemocratizingSustainability2023, Colesanti2024CombiningAA}. Recently, \citet{bulian2023assessing} developed a comprehensive evaluation framework based on science communication principles to assess the performance of LLMs in generating climate-related information.

\section{Data} \label{sec:data} 
\begin{figure*}[t]
	\centering
	\includegraphics[width=0.95\textwidth]{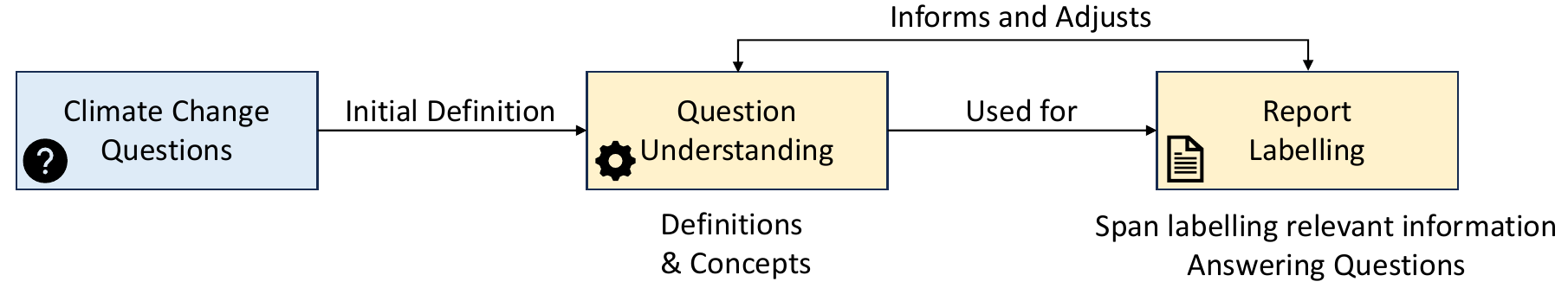}
	\caption{Labeling process to obtain the ClimRetrieve dataset.}
	\label{fig:labeling_process}
 \vspace{-1em}
\end{figure*}

This project constructs a dataset comprising authentic questions, sources, and answers to benchmark information retrieval in RAG systems in the use case of corporate climate disclosures. In this process, we simulate an analyst question-answering process based on documents.

The dataset creation involves an iterative question definition and report span labeling process (see Figure \ref{fig:labeling_process}). It starts with 16 Yes/No questions about climate change. The questions are inspired by the guidance of \citet{Bernhofen2023adaptation} and analyze companies' climate change adaptation. Thus, the question asks for details simulating an analyst's point of view on a company (see Appendix \ref{app:questions}). These questions are distributed among three expert annotators (see Appendix \ref{app:expert_annot}). For each question, an annotator creates a definition and concepts of the information sought in the question. Then, both are discussed in the expert group. This step is crucial to understanding the question in detail (see Appendix \ref{app:def_concepts} for details on the question definition and concepts).

In the next step, the expert annotators create the dataset using a specific sustainability report. Annotators search for relevant information in the report and annotate the sources from various perspectives. In this way, they replicate an analyst workflow in which the task is to read the document and search for relevant information to answer the question and rate its relevancy. Then, they answer the question based on the information. Ultimately, they create a dataset containing the following columns:

\begin{enumerate}[itemsep=0pt,topsep=1pt]
    \item \textbf{Document}: Report under investigation.
    \item \textbf{Question}: Question under investigation. 
    \item \textbf{Relevant}: Full-sentence form question-\textit{relevant information}.
    \item \textbf{Context}: Context of the question \textit{relevant information} (extending the relevant information by a sentence before and afterward).
    \item \textbf{Page}: Page of the \textit{relevant information}.
    \item \textbf{Source From}: Answers whether the \textit{relevant information} is from text, table or graph.
    \item \textbf{Source Relevance Score}: Classifies from 1-3 how relevant the information is for answering the question (see Appendix \ref{app:rel_labels} for details on the relevance classification). 
    \item \textbf{Unsure Flag}: Flag whether it is unclear if this source is question-relevant.
    \item \textbf{Addressed directly}: Flag whether the \textit{relevant information} addresses the question directly or indirectly.
    \item \textbf{Answer}: Answer to the question based on all retrieved \textit{relevant information}.
\end{enumerate}

After each report, the expert annotators have the option to discuss the question definitions and concepts with the expert group and retrofit them to the dataset. This allows for an iterative refinement of the nuances of question understanding.

This process is repeated for 30 sustainability reports. As a result, we obtain a base dataset with 743 entries of relevant question-source-answer pairs (see Appendix \ref{app:rel_qa_pairs} for details). Furthermore, we can create a report-level dataset since we know which parts of the report are relevant. In this dataset, we split the reports into paragraphs of equal length and mark relevant vs. nonrelevant parts with the question-source-answer pairs. This results in a dataset with 8.628 paragraphs labeled with the question's relevance. Since the questions are in semantic proximity, one paragraph can be relevant to multiple questions. For this reason, we ultimately create a dataset that contains unique report-paragraph-question pairs. For each question, the whole report is labeled. Thus, a report's paragraphs are repeated for each question to create an easy-to-assess dataset. In this way, we obtain a large report-level dataset with 43.445 entries (for details, see Appendix \ref{app:report_level_data}).


\section{Investigating Embedding Search}
We construct a specific use case to demonstrate the report-level dataset's practical applicability. Given the scarcity of research on information retrieval specific to climate-related corporate disclosures, this use case study is concentrated on this particular area.

Within the framework of a basic RAG model, inquiries posed to the document are utilized to identify pertinent paragraphs. This information retrieval typically follows a two-step process. First, embedding models are used to create a vector representation of the questions and all the paragraphs in the report. Second, the question vector is compared to all paragraph vectors to obtain the top k most similar paragraphs. However, as previous research has shown, LLMs are prone to be confused when presented with wrong or contradictory sources \citep{cuconasuPowerNoiseRedefining2024, watson2024hallucibot, schimanski2024faithful}, and the relevancy of the question to the sources plays a significant role \citep{niu2024ragtruth}. Thus, the retrieval process is central to creating the true output.

\begin{table*}
\centering
\begin{tabular}{lcccccc}
\hline
embeddings & question & definition & concepts & generic & inf\_3 & inf\_all \\
\hline
random & 0.037 & 0.037 & 0.037 & 0.037 & 0.037 & 0.037 \\
BM25 & 0.113 & 0.114 & 0.126 & 0.139 & 0.172 & \textbf{0.174} \\
ColBERTv2 & 0.109 & 0.094 & 0.112 & 0.124 & \textbf{0.137} & 0.130 \\
DRAGON+ & 0.139 & 0.121 & 0.106 & 0.141 & \textbf{0.161} & 0.160 \\
GTE-base & 0.161 & 0.153 & 0.171 & 0.154 & 0.171 & \textbf{0.174} \\
text-embedding-ada-002 & 0.163 & 0.140 & 0.157 & 0.155 & 0.178 & \textbf{0.179} \\
text-embedding-3-small & 0.163 & 0.143 & 0.140 & 0.161 & 0.175 & \textbf{0.176} \\
text-embedding-3-large & 0.167 & 0.143 & 0.152 & 0.163 & \textbf{0.179} & \textbf{0.179} \\
\hline
\end{tabular}
\caption{Table R.1: Results for the F1-score of the different embedding models and information retrieval approaches (question, definition, concepts, generic, inf\_3, inf\_all) aggregated across all top-k values (5, 10, 15). The best-performing information retrieval strategy (in bold) is the expert-informed explanations.}
\label{tab:base_results}
\end{table*}

As previously outlined, climate change is a complex downstream domain with knowledge-intensive questions (see Section \ref{sec:background} and Appendix \ref{app:def_concepts}). An expert labeler will likely consider additional concepts and definitions when searching for relevant information in reports. Thus, only using the question in the embedding search process might limit the results to semantically similar paragraphs to the question, not to all concepts embedded in the expert annotator's mind. 

Therefore, we construct an experiment that gradually replaces question ($question$) in the top-k search process with longer and more expert-informed question explanations. To obtain question explanations, we use two setups. First, we use the definitions and concepts the labelers used during their annotation (see Appendix \ref{app:def_concepts} for an example). Second, we make use of the capabilities of the closed-source LLM GPT-4. We proceed in two steps. In the first step, we ask the model to create an explanation for the question of (1) 60 words ($short$) and (2) 150 words ($long$). We further ask the model to include and exclude the question (e.g., $short\_Q$ / $short\_noQ$). These definitions serve as generic base cases ($generic$). In the second step, we gradually create more example-informed question explanations. In this artificial setup, we allow information leakage from the labeled \textit{relevant information} to inform the explanation creation process. We use the \textit{relevant information} with a label of 2 or higher as examples that should inspire the explanation (see Appendix \ref{app:rel_labels} for justification of the threshold). We create two settings: randomly choosing labeled \textit{relevant information} from three reports ($inf\_3$), and using all labeled \textit{relevant information} ($inf\_all$). For more details, see Appendix \ref{app:IR_queries}.

Finally, we employ simple evaluation metrics to compare the approaches. We define our first evaluation metric as the ratio of relevant sources found among all annotated sources. This equals the Recall@K. Thus, we try to optimize the number of relevant sources obtained by the information retrieval. Furthermore, we use the ratio of all relevant sources found in the retrieved sources, which equals the Precision@K in a classification task.

\begin{equation}
    Recall@K = \frac{relevant\_sources@K}{total\_relevant\_sources}
\end{equation}

\begin{equation}
    Precision@K =\frac{relevant\_sources@K}{K}
\end{equation}

This also allows us to calculate the weighted average, i.e., the F1 score. 
We calculate these scores at the top k values of 5, 10, and 15. Furthermore, we run the information retrieval with a variety of embedding models: a random baseline, BM25, DRAGON+, GTE-base, ColBERTv2, OpenAI's text-embedding-ada-002, text-embedding3-small, and text-embedding3-large (see Appendix \ref{app:experimental_setup} for details on the experimental setup). We analyze all setups individually as well as aggregate the scores over all embeddings or top-k values.

Our first step is to compare the questions in the retrieval process with the definitions and concepts written by the annotators. As Table \ref{tab:base_results} indicates, replacing the question with these definitions rather decreases the performance (see Appendix \ref{app:quest_def_concept_results} for more reinforcing results).

\begin{figure}[t]
    \centering	\includegraphics[width=0.49\textwidth]{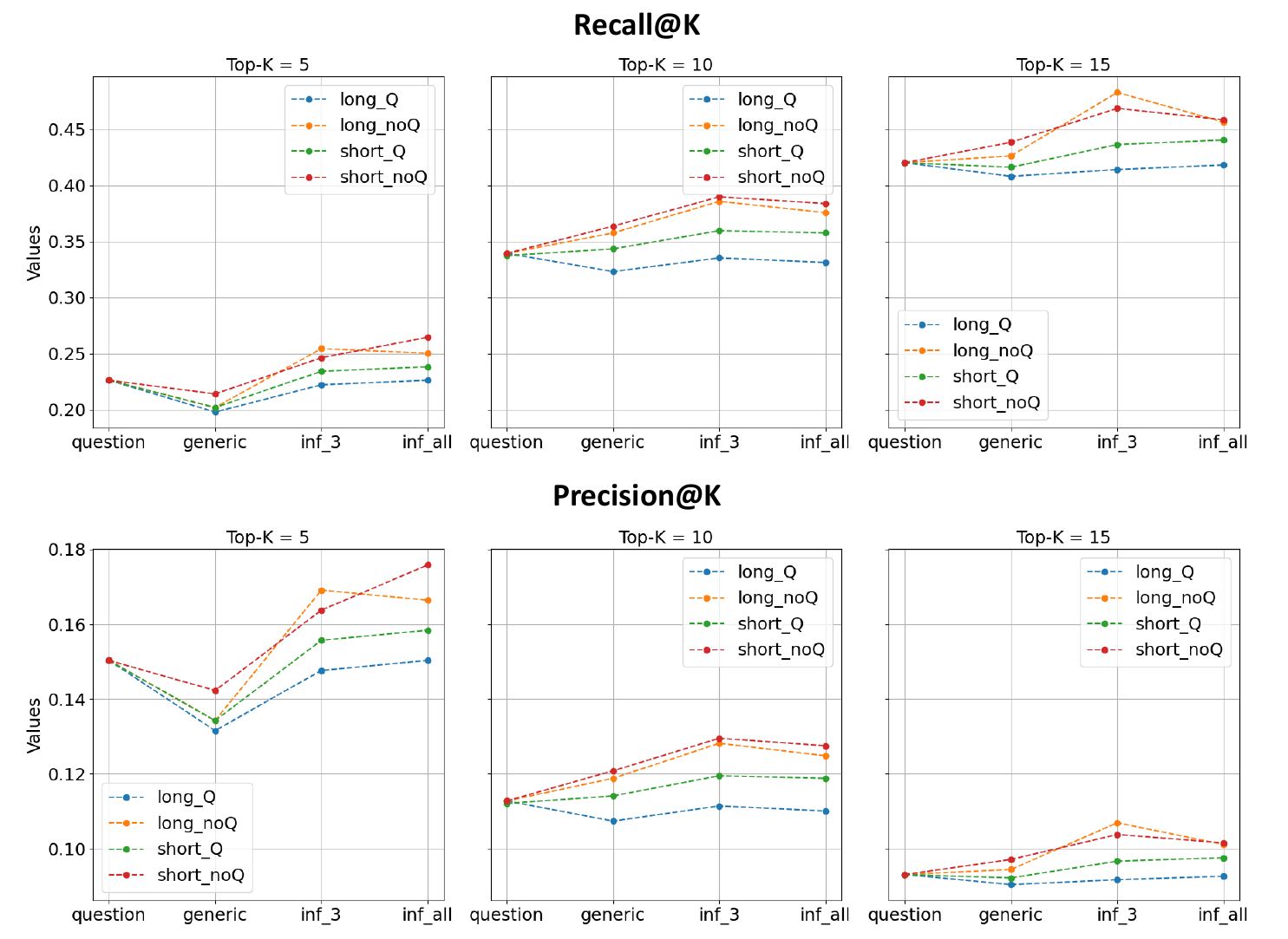}
	\caption{Results for the different experimental setups (Embeddings = "text-embedding-3-large").}
	\label{fig:compare_found}
 \vspace{-1em}
\end{figure}

This trend changes when using example-informed question explanations. As Table \ref{tab:base_results} shows generally, and Figure \ref{fig:compare_found} illustrates for just contemplating the results of text-embedding3-large, using these explanations can improve retrieval. The higher the top-k value, the more relevant sources are found in the retrieved ones. Also, the higher the top-k value, the less relevant sources are found relative to K (see Figure \ref{fig:compare_found}). Beyond these obvious insights, these results entail three major findings. First, using an example-informed, that is, an expert-informed explanation, improves the retrieval in contrast to using the definitions and concepts of the labelers. This probably originates from the fact that the example-inspired explanations offer greater detail tailored for the retrieval instead of capturing general concepts (see the appendix \ref{app:quest_def_concept_results} for comparison). Second, the most promising strategy for optimizing the embedding search is using expert-informed definitions that exclude the question. This is an interesting finding, indicating that the concept behind the questions seems to be more targeted for search than the question itself (results hold on aggregated scores, see Appendix \ref{app:Q_v_noQ}). Third, in light of the challenges around source quality and hallucination of LLMs, there is a need to improve efficient information retrieval processes. Although embeddings and using definitions certainly present a good first pathway, improvement in the nuance of question-source relevance beyond a fixed top-k number could improve the ultimate results. All these insights are consistently confirmed when considering different analysis metrics, embeddings, and relevance thresholds (see Appendix \ref{app:other_results} for these investigations).

\section{Conclusion} 
In this work, we develop a unique dataset that simulates an expert analyst workflow to evaluate RAG systems. We show its utility by analyzing the dominant embedding retrieval strategy with different search setups. We find that embeddings face major limitations in information retrieval for knowledge-intensive tasks. Therefore, this work sets the path for including and evaluating the improvement of expert-integrated information retrieval for RAG systems (see \citealp{ni2024dirasefficientllmassistedannotation} for a potential solution approach).

\section*{Limitations}
As with every work, our work has limitations. The first limitation comes from the expert workflow that we are using. Previous work has shown that experts face selection bias when annotating for information retrieval tasks \citep{thakurBEIRHeterogenousBenchmark2021}. This means that we certainly know that the source is relevant once labeled, but we do not know whether the source is irrelevant if not labeled. This likely means our results represent a lower bound rather than an absolute truth.

Second, as mentioned in creating the example-informed definitions, we intentionally allowed data leakage between the set to inspire the explanations and the test set. However, we argue that a real-world expert would act similarly when designing the explanations based on her previous experience.

\section*{Ethics Statement}
\myparagraph{Human Annotation}: In this work, all human annotators are Graduate or Doctorate researchers who have good knowledge about scientific communication and entailment. They are officially hired and have full knowledge of the context and utility of the collected data. We adhered strictly to ethical guidelines, respecting the dignity, rights, safety, and well-being of all participants.

\myparagraph{Data Privacy or Bias}: There are no data privacy issues or biases against certain demographics with regard to the data collected from real-world applications and LLM generations. All artifacts we use are under a Creative Commons license. We also notice no ethical risks associated with this work

\myparagraph{Reproducibility Statement}: To ensure full reproducibility, we will disclose all codes and data used in this project, as well as the LLM generations, GPT-4, and human annotations. For OpenAI models, we use ``gpt-4-0125-preview'' We always fix the temperature to 0 when using APIs.

\section*{Acknowledgements} This paper has received funding from the Swiss National Science Foundation (SNSF) under the project `How sustainable is sustainable finance? Impact evaluation and automated greenwashing detection' (Grant Agreement No. 100018\_207800).

\bibliography{custom}

\begin{thebibliography}{27}
\providecommand{\natexlab}[1]{#1}

\bibitem[{Asai et~al.(2023)Asai, Wu, Wang, Sil, and Hajishirzi}]{asai2023selfrag}
Akari Asai, Zeqiu Wu, Yizhong Wang, Avirup Sil, and Hannaneh Hajishirzi. 2023.
\newblock \href {https://arxiv.org/abs/2310.11511} {Self-rag: Learning to retrieve, generate, and critique through self-reflection}.
\newblock \emph{Preprint}, arXiv:2310.11511.

\bibitem[{Bernhofen and Ranger(2023)}]{Bernhofen2023adaptation}
Mark Bernhofen and Nicola Ranger. 2023.
\newblock Aligning finance with adaptation and resilience goals: Targets and metrics for financial institutions.
\newblock Technical report, University of Oxford, UK Center for Greening Finance \& Investment, Global Resilience Index Initiative.

\bibitem[{Bingler et~al.(2024)Bingler, Kraus, Leippold, and Webersinke}]{BINGLER2024107191}
Julia~Anna Bingler, Mathias Kraus, Markus Leippold, and Nicolas Webersinke. 2024.
\newblock \href {https://doi.org/10.1016/j.jbankfin.2024.107191} {How cheap talk in climate disclosures relates to climate initiatives, corporate emissions, and reputation risk}.
\newblock \emph{Journal of Banking \& Finance}, 164:107191.

\bibitem[{Bulian et~al.(2023)Bulian, Sch{\"a}fer, Amini, Lam, Ciaramita, Gaiarin, Huebscher, Buck, Mede, Leippold et~al.}]{bulian2023assessing}
Jannis Bulian, Mike~S Sch{\"a}fer, Afra Amini, Heidi Lam, Massimiliano Ciaramita, Ben Gaiarin, Michelle~Chen Huebscher, Christian Buck, Niels Mede, Markus Leippold, et~al. 2023.
\newblock Assessing large language models on climate information.
\newblock \emph{Proceedings of the ICML Conference, 2024, arXiv preprint arXiv:2310.02932}.

\bibitem[{Chen et~al.(2023)Chen, Lin, Han, and Sun}]{chenBenchmarkingLargeLanguage2023}
Jiawei Chen, Hongyu Lin, Xianpei Han, and Le~Sun. 2023.
\newblock \href {http://arxiv.org/abs/2309.01431} {Benchmarking {Large} {Language} {Models} in {Retrieval}-{Augmented} {Generation}}.
\newblock \emph{arXiv preprint}.
\newblock ArXiv:2309.01431 [cs].

\bibitem[{Colesanti~Senni et~al.(2024)Colesanti~Senni, Schimanski, Bingler, Ni, and Leippold}]{Colesanti2024CombiningAA}
Chiara Colesanti~Senni, Tobias Schimanski, Julia~Anna Bingler, Jingwei Ni, and Markus Leippold. 2024.
\newblock Combining ai and domain expertise to assess corporate climate transition disclosures.
\newblock \emph{SSRN Electronic Journal}.

\bibitem[{Commission(2024)}]{sec_rule}
Security~Exchange Commission. 2024.
\newblock \href {https://www.sec.gov/files/rules/final/2024/33-11275.pdf} {Final rule: The enhancement and standardization of climate-related disclosures for investors}.
\newblock Technical report, Securities and Exchange Commission (SEC).

\bibitem[{Cuconasu et~al.(2024)Cuconasu, Trappolini, Siciliano, Filice, Campagnano, Maarek, Tonellotto, and Silvestri}]{cuconasuPowerNoiseRedefining2024}
Florin Cuconasu, Giovanni Trappolini, Federico Siciliano, Simone Filice, Cesare Campagnano, Yoelle Maarek, Nicola Tonellotto, and Fabrizio Silvestri. 2024.
\newblock \href {http://arxiv.org/abs/2401.14887} {The {Power} of {Noise}: {Redefining} {Retrieval} for {RAG} {Systems}}.
\newblock \emph{arXiv preprint}.
\newblock ArXiv:2401.14887 [cs].

\bibitem[{Diggelmann et~al.(2021)Diggelmann, Boyd-Graber, Bulian, Ciaramita, and Leippold}]{diggelmann2021climatefever}
Thomas Diggelmann, Jordan Boyd-Graber, Jannis Bulian, Massimiliano Ciaramita, and Markus Leippold. 2021.
\newblock Climate-fever: A dataset for verification of real-world climate claims.
\newblock \emph{arXiv preprint arXiv:2012.00614}.

\bibitem[{Kölbel et~al.(2022)Kölbel, Leippold, Rillaerts, and Wang}]{koelbel22ask}
Julian~F Kölbel, Markus Leippold, Jordy Rillaerts, and Qian Wang. 2022.
\newblock {Ask BERT: How Regulatory Disclosure of Transition and Physical Climate Risks Affects the CDS Term Structure*}.
\newblock \emph{Journal of Financial Econometrics}.

\bibitem[{Leippold et~al.(2024)Leippold, Vaghefi, Stammbach, Muccione, Bingler, Ni, Colesanti-Senni, Wekhof, Schimanski, Gostlow, Yu, Luterbacher, and Huggel}]{leippold2024automated}
Markus Leippold, Saeid~Ashraf Vaghefi, Dominik Stammbach, Veruska Muccione, Julia Bingler, Jingwei Ni, Chiara Colesanti-Senni, Tobias Wekhof, Tobias Schimanski, Glen Gostlow, Tingyu Yu, Juerg Luterbacher, and Christian Huggel. 2024.
\newblock \href {https://arxiv.org/abs/2401.12566} {Automated fact-checking of climate change claims with large language models}.
\newblock \emph{Preprint}, arXiv:2401.12566.

\bibitem[{Lewis et~al.(2020)Lewis, Perez, Piktus, Petroni, Karpukhin, Goyal, K\"{u}ttler, Lewis, Yih, Rockt\"{a}schel, Riedel, and Kiela}]{Lewis_2020}
Patrick Lewis, Ethan Perez, Aleksandra Piktus, Fabio Petroni, Vladimir Karpukhin, Naman Goyal, Heinrich K\"{u}ttler, Mike Lewis, Wen-tau Yih, Tim Rockt\"{a}schel, Sebastian Riedel, and Douwe Kiela. 2020.
\newblock Retrieval-augmented generation for knowledge-intensive nlp tasks.
\newblock In \emph{Advances in Neural Information Processing Systems}, volume~33, pages 9459--9474. Curran Associates, Inc.

\bibitem[{Ni et~al.(2023)Ni, Bingler, Colesanti-Senni, Kraus, Gostlow, Schimanski, Stammbach, Ashraf~Vaghefi, Wang, Webersinke, Wekhof, Yu, and Leippold}]{niCHATREPORTDemocratizingSustainability2023}
Jingwei Ni, Julia Bingler, Chiara Colesanti-Senni, Mathias Kraus, Glen Gostlow, Tobias Schimanski, Dominik Stammbach, Saeid Ashraf~Vaghefi, Qian Wang, Nicolas Webersinke, Tobias Wekhof, Tingyu Yu, and Markus Leippold. 2023.
\newblock \href {https://doi.org/10.18653/v1/2023.emnlp-demo.3} {{CHATREPORT}: {Democratizing} {Sustainability} {Disclosure} {Analysis} through {LLM}-based {Tools}}.
\newblock In \emph{Proceedings of the 2023 {Conference} on {Empirical} {Methods} in {Natural} {Language} {Processing}: {System} {Demonstrations}}, pages 21--51, Singapore. Association for Computational Linguistics.

\bibitem[{Ni et~al.(2024)Ni, Schimanski, Lin, Sachan, Ash, and Leippold}]{ni2024dirasefficientllmassistedannotation}
Jingwei Ni, Tobias Schimanski, Meihong Lin, Mrinmaya Sachan, Elliott Ash, and Markus Leippold. 2024.
\newblock \href {https://arxiv.org/abs/2406.14162} {Diras: Efficient llm-assisted annotation of document relevance in retrieval augmented generation}.
\newblock \emph{Preprint}, arXiv:2406.14162.

\bibitem[{Niu et~al.(2024)Niu, Wu, Zhu, Xu, Shum, Zhong, Song, and Zhang}]{niu2024ragtruth}
Cheng Niu, Yuanhao Wu, Juno Zhu, Siliang Xu, Kashun Shum, Randy Zhong, Juntong Song, and Tong Zhang. 2024.
\newblock \href {https://arxiv.org/abs/2401.00396} {Ragtruth: A hallucination corpus for developing trustworthy retrieval-augmented language models}.
\newblock \emph{Preprint}, arXiv:2401.00396.

\bibitem[{Saad-Falcon et~al.(2024)Saad-Falcon, Khattab, Potts, and Zaharia}]{saadfalcon2024ares}
Jon Saad-Falcon, Omar Khattab, Christopher Potts, and Matei Zaharia. 2024.
\newblock \href {https://arxiv.org/abs/2311.09476} {Ares: An automated evaluation framework for retrieval-augmented generation systems}.
\newblock \emph{Preprint}, arXiv:2311.09476.

\bibitem[{Schimanski et~al.(2023)Schimanski, Bingler, Kraus, Hyslop, and Leippold}]{schimanski-etal-2023-climatebert}
Tobias Schimanski, Julia Bingler, Mathias Kraus, Camilla Hyslop, and Markus Leippold. 2023.
\newblock \href {https://doi.org/10.18653/v1/2023.emnlp-main.975} {{C}limate{BERT}-{N}et{Z}ero: Detecting and assessing net zero and reduction targets}.
\newblock In \emph{Proceedings of the 2023 Conference on Empirical Methods in Natural Language Processing}, pages 15745--15756, Singapore. Association for Computational Linguistics.

\bibitem[{Schimanski et~al.(2024{\natexlab{a}})Schimanski, Ni, Kraus, Ash, and Leippold}]{schimanski2024faithful}
Tobias Schimanski, Jingwei Ni, Mathias Kraus, Elliott Ash, and Markus Leippold. 2024{\natexlab{a}}.
\newblock \href {https://arxiv.org/abs/2402.08277} {Towards faithful and robust llm specialists for evidence-based question-answering}.
\newblock \emph{Preprint}, arXiv:2402.08277.

\bibitem[{Schimanski et~al.(2024{\natexlab{b}})Schimanski, Reding, Reding, Bingler, Kraus, and Leippold}]{SCHIMANSKI2024104979}
Tobias Schimanski, Andrin Reding, Nico Reding, Julia Bingler, Mathias Kraus, and Markus Leippold. 2024{\natexlab{b}}.
\newblock \href {https://doi.org/10.1016/j.frl.2024.104979} {Bridging the gap in esg measurement: Using nlp to quantify environmental, social, and governance communication}.
\newblock \emph{Finance Research Letters}, 61:104979.

\bibitem[{Sietsma et~al.(2023)Sietsma, Ford, and Minx}]{Sietsma2023TheNG}
Anne~J. Sietsma, James~D. Ford, and Jan~C. Minx. 2023.
\newblock The next generation of machine learning for tracking adaptation texts.
\newblock \emph{Nature Climate Change}.

\bibitem[{Stammbach et~al.(2023)Stammbach, Webersinke, Bingler, Kraus, and Leippold}]{stammbach2022dataset}
Dominik Stammbach, Nicolas Webersinke, Julia~Anna Bingler, Mathias Kraus, and Markus Leippold. 2023.
\newblock Environmental claim detection.
\newblock In \emph{Proceedings of the 61st Annual Meeting of the Association for Computational Linguistics}, pages 1051--1066. Association for Computational Linguistics.

\bibitem[{Thakur et~al.(2021)Thakur, Reimers, Rücklé, Srivastava, and Gurevych}]{thakurBEIRHeterogenousBenchmark2021}
Nandan Thakur, Nils Reimers, Andreas Rücklé, Abhishek Srivastava, and Iryna Gurevych. 2021.
\newblock \href {http://arxiv.org/abs/2104.08663} {{BEIR}: {A} {Heterogenous} {Benchmark} for {Zero}-shot {Evaluation} of {Information} {Retrieval} {Models}}.
\newblock \emph{arXiv preprint}.
\newblock ArXiv:2104.08663 [cs].

\bibitem[{Vaghefi et~al.(2023)Vaghefi, Stammbach, Muccione, Bingler, Ni, Kraus, Allen, Colesanti-Senni, Wekhof, Schimanski, Gostlow, Yu, Wang, Webersinke, Huggel, and Leippold}]{vaghefi2023chatclimate}
Saeid~Ashraf Vaghefi, Dominik Stammbach, Veruska Muccione, Julia Bingler, Jingwei Ni, Mathias Kraus, Simon Allen, Chiara Colesanti-Senni, Tobias Wekhof, Tobias Schimanski, Glen Gostlow, Tingyu Yu, Qian Wang, Nicolas Webersinke, Christian Huggel, and Markus Leippold. 2023.
\newblock \href {https://doi.org/10.1038/s43247-023-01084-x} {Chatclimate: Grounding conversational ai in climate science}.
\newblock \emph{Communications Earth \& Environment}, 4(1):480.

\bibitem[{Varini et~al.(2021)Varini, Boyd-Graber, Ciaramita, and Leippold}]{varini2021climatext}
Francesco~S. Varini, Jordan Boyd-Graber, Massimiliano Ciaramita, and Markus Leippold. 2021.
\newblock Climatext: A dataset for climate change topic detection.
\newblock \emph{arXiv preprint arXiv:2012.00483}.

\bibitem[{Watson and Cho(2024)}]{watson2024hallucibot}
William Watson and Nicole Cho. 2024.
\newblock \href {https://arxiv.org/abs/2404.12535} {Hallucibot: Is there no such thing as a bad question?}
\newblock \emph{Preprint}, arXiv:2404.12535.

\bibitem[{Weber and Baisch(2023)}]{Weber2023ClimateCR}
Rolf~H. Weber and Rainer Baisch. 2023.
\newblock \href {https://api.semanticscholar.org/CorpusID:257819558} {Climate change reporting and human information processing – quo vadis transparency?}
\newblock \emph{ex/ante}.

\bibitem[{Zhang et~al.(2024)Zhang, Patil, Jain, Shen, Zaharia, Stoica, and Gonzalez}]{zhang2024raft}
Tianjun Zhang, Shishir~G. Patil, Naman Jain, Sheng Shen, Matei Zaharia, Ion Stoica, and Joseph~E. Gonzalez. 2024.
\newblock \href {https://arxiv.org/abs/2403.10131} {Raft: Adapting language model to domain specific rag}.
\newblock \emph{Preprint}, arXiv:2403.10131.

\end{thebibliography}

\appendix

\section*{Appendix}

\setcounter{section}{0}  
\renewcommand\thefigure{\thesection.\arabic{figure}}   
\setcounter{figure}{0}   
\renewcommand\thetable{\thesection.\arabic{table}}   
\setcounter{table}{0}   

\section{Complexity of Knowledge-Intensive Questions}\label{app:knowledge_intensiveQ}
Knowledge-intensive domains like climate change have knowledge-intensive questions. Consider, i.e., the following question: "What are the company's emissions for the previous year?". While emissions serve as a fundamental indicator of a company's environmental impact, the associated complexities are profound. Emissions can be stratified into various categories, including carbon dioxide (CO2), methane (CH4), among others. Moreover, it is increasingly critical to distinguish between direct, indirect, and supply chain emissions, both upstream and downstream (Scope 1-3). This example underscores the extensive complexity that must be integrated into the analysis of ostensibly straightforward questions.

\section{Definitions and Concepts}\label{app:def_concepts}
For knowledge-intensive domains like climate change, it is of central importance to obtain the right question understanding. As demonstrated with the emission example in Appendix \ref{app:knowledge_intensiveQ}, simple questions can unfold a large underlying mass of concepts. 

Generally, two differentiations are used in this work. When an expert reads a question, she might have two things in mind: definitions and concepts. On the one hand, definitions constitute the elucidation of the terminologies referenced within the query. For example, when inquiring about emissions, one might interpret them as the gases the company generates during its value-creation processes. This interpretation is inherently complex and varies significantly among experts and specific use cases. 

Conversely, concepts pertain to the interconnected themes associated with the questions. We can distinguish between two types of concepts. First, core concepts are intrinsically linked to the query and exhibit substantial overlap with definitions. For example, in the question "What are the emissions of the company in the last year?", the term "emissions" constitutes a core concept. However, the phrase "last year" introduces potential ambiguity if not explicitly defined—whether it refers to a reporting year or a calendar year. Second, lateral concepts represent broader, knowledge-graph-like connections. For instance, in the context of emissions, a lateral concept might encompass climate change. An expert's interpretation of the lateral concepts in the question "What are the emissions of the company in the last year?" could extend to inquiries regarding climate change mitigation. Given these considerations, it is imperative to elucidate both definitions and concepts when seeking information and formulating responses. 

These concepts and definitions could manifest entirely differently depending from person to person. For this dataset, the important thing is that the question sources, answers, definitions, and concepts are consistent with itself. Table \ref{tab:concepts_defs} gives an example of a definition and concepts for the question ""Do the environmental/sustainability targets set by the company reference external climate change adaptation goals/targets?"".\footnote{All data and code is open-source under \url{https://github.com/tobischimanski/ClimRetrieve}.}

\begin{table*}[]
\begin{tabular}{p{3cm}p{5cm}p{7cm}}
\hline
Question                                                                                                                   & Definition                                                                                                                                                                                                                                                                                                                                                                  & Concepts                                                                                                                                                                                                                                                                                                                                                                                                                                                                                                                                                                                                                                                                                                                                                                                                                                                                                                                                                                                                                                                                                                                                                                                                                                                                                                                                                                                                                                                                                                                                                                                                                                                                                                        \\ \hline
Do the environmental/sustainability targets set by the company reference external climate change adaptation goals/targets? & \begin{lstlisting}[frame=single, basicstyle=\ttfamily\scriptsize,  breaklines=true , xleftmargin=0pt, numbers=none, label=code:prompt_1]
External climate change adaptation goals or targets include national, regional or sectoral adaptation plans set either by government, industry bodies, standard setters, or international organisations such as the United Nations, the World Bank or others. The external targets must be provided. 
\end{lstlisting} & 
\begin{lstlisting}[frame=single, basicstyle=\ttfamily\scriptsize,  breaklines=true , xleftmargin=0pt, numbers=none, label=code:prompt_1]
1. [Core] **Reducing Greenhouse Gas Emissions**: Setting targets to decrease emissions of carbon dioxide (CO2), methane (CH4), and other greenhouse gases to mitigate climate change.
2. [Core] **Increasing Renewable Energy Usage**: Establishing goals to increase the percentage of energy generated from renewable sources such as solar, wind, hydroelectric, and geothermal power.
3. [Latent] **Conservation of Biodiversity**: Setting targets to preserve and protect natural habitats, endangered species, and ecosystems to maintain biodiversity.
4. [Latent] **Reducing Waste and Promoting Recycling**: Implementing measures to minimize waste generation, increase recycling rates, and promote a circular economy.
5. [Latent] **Water Management and Conservation**: Developing strategies to manage water resources more efficiently, such as investing in water-saving technologies, implementing rainwater harvesting systems, and improving water storage and distribution infrastructure to cope with changing precipitation patterns and droughts
6. [Core] **Building Climate-Resilient Infrastructure**: Integrating climate resilience into infrastructure planning and design, including constructing buildings and roads that can withstand extreme weather events, improving drainage systems to manage flooding, and upgrading energy and transportation networks to reduce vulnerability to climate impacts.
7. [Core] **Enhancing Disaster Preparedness and Response**: Developing early warning systems, emergency response plans, and community resilience programs to prepare for and respond to natural disasters such as hurricanes, floods, wildfires, and heatwaves.
\end{lstlisting}
\\ \hline
\end{tabular}
\caption{Example of a question definition and concepts the labler articulates about the question. Concepts are differentiate to be [Core] or [Latent] Concepts.}
\label{tab:concepts_defs}
\end{table*}

\section{Questions}\label{app:questions}
Table \ref{tab:questions} displays the questions the expert annotators answered for the reports. The focus lies on climate change adaptation and the resilience of companies. Thus, the questions are detailed and specific. The questions were created based on the guidance by \citet{Bernhofen2023adaptation}. Furthermore, all questions are designed to be answerable with Yes or No and a free text explanation. This offers a nuanced level of detail in the potential analyses. In this project, we focus on the retrieved sources and not on the answers because retrieval is much less researched and the source dataset offers a richer amount of analysis potential.

\begin{table*}[]
\begin{tabular}{p{0.5cm}p{15cm}}
\hline
   & \textbf{Question}                                                                                                                                           \\ \hline
1  & Does the company have a specific process in place to identify risks arising from climate change?                                                            \\
2  & Does the company report the methodology used to identify the dependencies and impact of its business activities on the environment?                         \\
3  & Does the company refer to any third party scenarios when identifying climate-related risks or opportunities (e.g. IPCC trajectories, NGFS scenarios, etc.)? \\
4  & Does the company encourage downstream partners to carry out climate-related risk assessments?                                                               \\
5  & Does the company report how adjustments to its business operations will allow it to adapt to climate change?                                                \\
6  & Does the company provide definitions for climate change adaptation?                                                                                         \\
7  & Has the company identified any synergies between its climate change adaptation goals and other business goals?                                              \\
8  & Does the company report the climate change scenarios used to test the resilience of its business strategy?                                                  \\
9  & Does the company seek to adjust its business model to better provide climate change adaptation products and services?                                       \\
10 & Does the company have any engagements with industry peers in relation to climate change?                                                                    \\
11 & Do the environmental/sustainability targets set by the company reference external climate change adaptation goals/targets?                                  \\
12 & Do the environmental/sustainability targets set by the company align with external climate change adaptation goals/targets?                                 \\
13 & Does the company report short-term actions taken or planned to reduce its waste generation?                                                                 \\
14 & Does the company report a plan to engage with downstream partners on water consumption or water pollution?                                                  \\
15 & Does the company identify any impacts of its business activities on the environment?                                                                        \\
16 & Does the company have a strategy on waste management?                                                                                                       \\ \hline
\end{tabular}
\caption{Questions the expert annotators labeled for the reports.}
\label{tab:questions}
\end{table*}

\section{Expert Annotators and Expert Group}\label{app:expert_annot}
The three annotators involved in this study hold an undergraduate degree with a minor or major focus in the climate domain. All annotators have at least one year of professional experience in the field. During the process of labeling, all annotators are enrolled in a master's program with a focus in the sustainability or climate domain at the University of Oxford.

The expert group in this project is composed of the three expert annotators, two junior and one senior researcher in the domain. The expert group collectively defined questions, discussed definitions and concepts for the questions and was involved in the iterative refinement of the dataset. 

\section{Relevance Labels of the Dataset}\label{app:rel_labels}
For answering a question, texts of different relevance can be in a report. To reflect this fact, we introduce three relevance labels where 1 is partially relevant, 2 is relevant, and 3 is highly relevant. This means, there is a clear difference between 2 and 3 being certainly relevant and 1 where the labeler might be unsure about relevance or can only identify indirect relevance. However, this also means that experiments using the final dataset may want to reflect the fact that a paragraph with label 1 differs from those with labels 2 and 3.

\section{Relevant Question-Source-Answer Pairs}\label{app:rel_qa_pairs}
The core result of the labeling process is 743 question-source-answer pairs with the 16 questions under consideration. For each question, sources are searched, labeled by relevance and the other categories (see Section \ref{sec:data}), and finally answered. The questions are split amongst the annotators so that two annotators label 5 questions per report and one annotator labels 6 questions per report. As Table \ref{tab:qsa-pairs} shows, there is a discrepancy in how many question-source-answer pairs per question exist in the dataset. The determining factor for this variance is the number of sources found per question. While more sources can be found for more general questions like ""Does the company have a specific process in place to identify risks arising from climate change?"" (66 sources found across the dataset), detailed questions like ""Does the company provide definitions for climate change adaptation?"" are less often answered through the reports (6 sources found across the dataset). Thus, the dataset also contains questions where no sources were found.

\begin{table*}[]
\centering
\begin{tabular}{cccccccl}
\hline
count & mean & std & min & 25\% & 50\% & 75\% & max \\ \hline
16.0    & 37.2  & 17.6 & 13.0  & 27.7 & 34.0   & 48.0  & 72.0 \\ \hline
\end{tabular}
\caption{Descriptive statistics of the question-source-answer pairs per question.}
\label{tab:qsa-pairs}
\end{table*}

\begin{figure}[t]
    \centering	\includegraphics[width=0.48\textwidth]{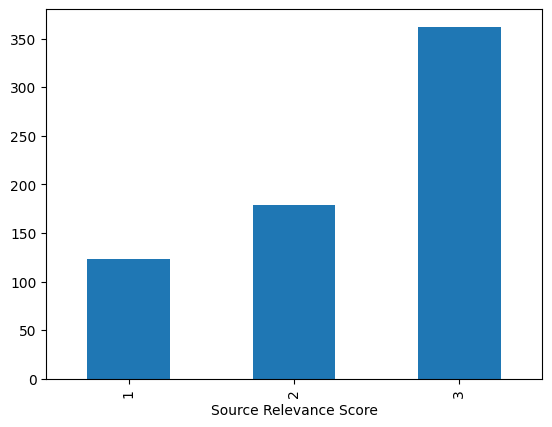}
	\caption{Distribution of relevance labels over the relevant question-source-answer dataset.}
	\label{fig:Dist_Source_Relevance_Score}
\end{figure}

After labeling, we arrive at a dataset containing majorly relevant question-source-answer. As Figure \ref{fig:Dist_Source_Relevance_Score} shows, the majority of the relevant question-source-answer pairs are indeed very relevant (relevance label 3). This speaks for the nature of the analyst workflow employed in this work where an analyst will likely search for the most relevant bits of information to answer the question.

\section{Report-Level Dataset}\label{app:report_level_data}
To obtain a report-level dataset of relevant vs. nonrelevant paragraphs, we use the LLamaIndex SentenceSplitter function.\footnote{See \url{https://docs.llamaindex.ai/en/stable/api_reference/node_parsers/sentence_splitter/} for more details.} This function allows the splitting of a document around a fixed length but tries to ensure the full-sentence form of the paragraphs. We specify the paragraph length to be around 350 words, while we allow for an overlap in paragraphs of 50 words. The overlap should prevent the loss of context through random cut-offs. This results in obtaining a dataset with 8628 paragraphs from the 30 reports.

\begin{figure}[t]
    \centering	\includegraphics[width=0.48\textwidth]{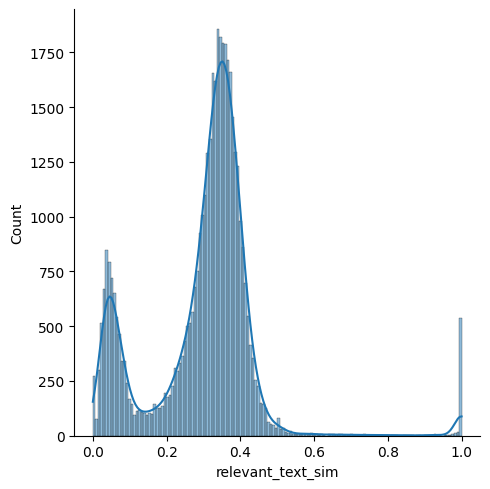}
	\caption{Similarities of the most similar relevant text part from the question-source-answer pairs with the paragraphs from the report-level dataset.}
	\label{fig:relevant_text_sim_dist}
\end{figure}

Once we obtain the paragraphs, we use our dataset with relevant question-source-answer pairs to assign a label to the whole set of paragraphs. Since the annotated dataset contains relevant sentences, we deem a paragraph relevant once it contains one of the sentences of the relevant text parts. The retrieved paragraphs from the reports sometimes have minor differences from the ones in the dataset, e.g. different spacing or headlines are included by the SentenceSplitter function. Thus, we use the difflab SequenceMatcher function to compare the similarity of sentences.\footnote{See \url{https://docs.python.org/3/library/difflib.html} for more details.} We use a similarity threshold of 0.9 for matching. This is orientated on experimentation with examples. However, the majority of the samples are clearly matchable with this threshold. Figure \ref{fig:relevant_text_sim_dist} shows the similarities between the most similar relevant text part from the question-source-answer pairs with the paragraphs from the report-level dataset. It becomes apparent that the paragraphs are either extremely similar to the sources (i.e., it's a match) or very dissimilar indicating that there is indeed no match found.

Since we want to obtain a dataset where every paragraph obtains a relevance score toward a question, we have to repeat the matching for each question that was answered for the report. Thus, we obtain a dataset with 43.445 entries from the original 8.628. These paragraphs now can appear multiple times with multiple questions. In its essence, the final report-level dataset contains pairs of paragraphs with questions. For each question, a relevance label is given between 0 (no relevance) and 1-3 (labeled as relevant by annotators). If the paragraph is relevant, we also give the relevant text part with which it was matched.

We fail to match the entire 743 question-source-answer pairs with the report-level dataset. This originates from problems with the chunking of the reports (e.g., not every paragraph is parsed correctly), issues when matching (e.g., the string was formatted differently and the threshold was not low enough), or the fact the information is retrieved from graphs or tables where the string matching doesn't work either. Finally, the report-level dataset contains 595 paragraphs with question-relevant information. Some paragraphs are relevant for multiple questions. The number of relevant unique paragraphs is 446 (within the 8.628).

\section{Information Retrieval Explanation}\label{app:IR_queries}
To replace the questions in the information retrieval process with definitions, we create generic and example-inspired explanations. Generic explanations simply take the question and create an explanation with the embedded knowledge of GPT-4 (the gpt-4-0125-preview checkpoint is used for all generations). We differentiate between explanations with the question (see Prompt \ref{lst:prompt_base_case_Q}) and without question (see Prompt \ref{lst:prompt_base_case_noQ}). This serves as a non-informed base case. To inform the question with actually relevant content, we make use of the already labeled relevant paragraphs and ask the model to abstract from these examples to create informed explanations. We again create an explanation with and without question (see Prompt \ref{lst:prompt_spec_case_Q} and Prompt \ref{lst:prompt_spec_case_noQ}). In the labeled dataset, the sources' relevance is differentiated from 1 (loosely relevant) to 3 (highly relevant). In order to ensure that only specific information informs the explanation creation process, we only consider sources of relevance 2 and higher as examples. We create explanations of different lengths (60 and 150 words) and with and without the questions. To illustrate these explanations, refer to Table \ref{tab:query_gen_inf3_60} with examples of length 60 without question and Table \ref{tab:query_gen_infall_150} with examples of length 150 with the question. While the beginning of the query remains the same, the longer queries might have different shapes in terms of containing lists or enumerations.

\begin{table*}[]
\begin{tabular}{p{3cm}p{6cm}p{6cm}}
\hline
Question                                                                                                                   & Generic Explanation                                                                                                                                                                                                                                                                                                                                                                  & Explanation Inspired by Three Reports                                          \\ \hline
Do the environmental/sustainability targets set by the company reference external climate change adaptation goals/targets? & \begin{lstlisting}[frame=single, basicstyle=\ttfamily\scriptsize,  breaklines=true , xleftmargin=0pt, numbers=none, label=code:prompt_1]
We search for details on whether the company's sustainability objectives align with broader climate change adaptation benchmarks, such as those outlined by international agreements (e.g., Paris Agreement) or national adaptation plans. This includes examining if goals address enhancing resilience to climate impacts, integrating climate adaptation into business strategies, and contributing to global efforts to adapt to changing climate conditions. 
\end{lstlisting} & 
\begin{lstlisting}[frame=single, basicstyle=\ttfamily\scriptsize,  breaklines=true , xleftmargin=0pt, numbers=none, label=code:prompt_1]
We search for details on how a company's sustainability goals align with recognized external climate change frameworks or initiatives, such as the UN's early warning systems, the Science Based Targets initiative, the Paris Agreement, or the ISO Net Zero Guidelines. This includes commitments to renewable energy, emissions reduction, and investments in nature-based solutions, demonstrating alignment with global efforts to combat climate change and promote resilience.
\end{lstlisting}
\\ \hline
\end{tabular}
\caption{Example of information retrieval explanations of length of 60 words excluding the question.}
\label{tab:query_gen_inf3_60}
\end{table*}

\begin{table*}[]
\begin{tabular}{p{3cm}p{6cm}p{6cm}}
\hline
Question                                                                                                                   & Generic Explanation                                                                                                                                                                                                                                                                                                                                                                  & Explanation Inspired by Three Reports                                          \\ \hline
Do the environmental/sustainability targets set by the company reference external climate change adaptation goals/targets? & \begin{lstlisting}[frame=single, basicstyle=\ttfamily\scriptsize,  breaklines=true , xleftmargin=0pt, numbers=none, label=code:prompt_1]
The question ""Do the environmental/sustainability targets set by the company reference external climate change adaptation goals/targets?"" is asking for details on whether the company's sustainability or environmental objectives align with broader, externally established climate change adaptation and resilience benchmarks or goals. This includes understanding if the company has integrated international, national, or sector-specific adaptation strategies into their sustainability planning. 

Examples of information the analyst is looking for include:
- Mention of adherence to frameworks like the Paris Agreement, the United Nations Sustainable Development Goals (SDGs), particularly SDG 13 (Climate Action), or the Sendai Framework for Disaster Risk Reduction.
- References to national adaptation plans or strategies that the company has aligned with.
- Inclusion of sector-specific resilience standards or benchmarks in the company's sustainability targets.
- Partnerships or collaborations with external bodies focused on climate change adaptation and resilience.
- Specific adaptation measures or targets that address identified climate risks relevant to the company's operations or value chain.\end{lstlisting} & 
\begin{lstlisting}[frame=single, basicstyle=\ttfamily\scriptsize,  breaklines=true , xleftmargin=0pt, numbers=none, label=code:prompt_1]
The question "Do the environmental/sustainability targets set by the company reference external climate change adaptation goals/targets?" is asking for details on how a company's sustainability or environmental objectives align with broader, recognized climate change adaptation and resilience frameworks or initiatives. This includes looking for evidence that the company has set its environmental targets in response to or in alignment with international agreements (such as the Paris Agreement), initiatives by global organizations (like the UN or the Science Based Targets initiative), or standards and guidelines set by authoritative bodies (such as the International Organization for Standardization). The question seeks to identify whether the company is not only setting internal goals but also contributing to global efforts to combat climate change through adaptation and resilience. This could involve commitments to renewable energy, science-based targets for reducing greenhouse gas emissions, investments in nature-based solutions, or participation in global calls to action for climate resilience. The aim is to gauge the company's active engagement in the global climate adaptation agenda beyond its immediate operational boundaries.
\end{lstlisting}
\\ \hline
\end{tabular}
\caption{Example of information retrieval explanations of length 150 words including the question.}
\label{tab:query_gen_infall_150}
\end{table*}

\begin{figure*}[ht]
\begin{lstlisting}[frame=single, basicstyle=\ttfamily\scriptsize,  breaklines=true , xleftmargin=0pt, numbers=none, label=code:prompt_1]
You are a sustainability report analyst specialising on climate change adaptation and resilience.

You are provided with a <QUESTION> about a sustainability report. Your task is to explain the <QUESTION> in the context of adaptation and resilience. Please first explain the meaning of the <question>, i.e., meaning of the question itself and the concepts mentioned. And then give a list of examples, showing what information from the sustainability report the analyst is looking for by posting this <question>.

The <QUESTION> is:
{question}

Your task is to create a short {length} word explanation for which details the question is asking for.

Start the answer with 'The question "<QUESTION>" is asking for details on...'.

Your answer:
\end{lstlisting}
\caption{\label{lst:prompt_base_case_Q} Prompt for creating the generic information retrieval explanation \textbf{with} the question.}
\end{figure*}

\begin{figure*}[ht]
\begin{lstlisting}[frame=single, basicstyle=\ttfamily\scriptsize,  breaklines=true , xleftmargin=0pt, numbers=none, label=code:prompt_1]
You are a sustainability report analyst specialising on climate change adaptation and resilience.

You are provided with a <QUESTION> about a sustainability report. Your task is to explain the <QUESTION> in the context of adaptation and resilience. Please first explain the meaning of the <question>, i.e., meaning of the question itself and the concepts mentioned. And then give a list of examples, showing what information from the sustainability report the analyst is looking for by posting this <question>.

The <QUESTION> is:
{0}

Your task is to create a short {1} word explanation for which details the question is asking for.

Start the answer with 'We search for details on'. Don't mention the question itself in the text.
Your answer:
\end{lstlisting}
\caption{\label{lst:prompt_base_case_noQ} Prompt for creating the generic information retrieval explanation \textbf{without} the question.}
\end{figure*}

\begin{figure*}[ht]
\begin{lstlisting}[frame=single, basicstyle=\ttfamily\scriptsize,  breaklines=true , xleftmargin=0pt, numbers=none, label=code:prompt_1]
You are a sustainability report analyst specialising on climate change adaptation and resilience.

You are provided with a <QUESTION> about a sustainability report. Your task is to explain the <QUESTION> in the context of adaptation and resilience. Please first explain the meaning of the <question>, i.e., meaning of the question itself and the concepts mentioned. And then give a list of examples, showing what information from the sustainability report the analyst is looking for by posting this <question>.

The <QUESTION> is:
{question}

Furthermore, you already analysed reports and extracted the following passages of relevant information the question is looking for:
---
{examples}
---

Your task is to create a short {length} word explanation for which details the question is asking for.
Make sure to make use of the passages by not directly referencing them but using them to influence the details that might be of help.

Start the answer with 'The question "<QUESTION>" is asking for details on...'.

Your answer:
\end{lstlisting}
\caption{\label{lst:prompt_spec_case_Q} Prompt for creating the expert-informed information retrieval explanation \textbf{with} the question.}
\end{figure*}

\begin{figure*}[ht]
\begin{lstlisting}[frame=single, basicstyle=\ttfamily\scriptsize,  breaklines=true , xleftmargin=0pt, numbers=none, label=code:prompt_1]
You are a sustainability report analyst specialising on climate change adaptation and resilience.

You are provided with a <QUESTION> about a sustainability report. Your task is to explain the <QUESTION> in the context of adaptation and resilience. Please first explain the meaning of the <question>, i.e., meaning of the question itself and the concepts mentioned. And then give a list of examples, showing what information from the sustainability report the analyst is looking for by posting this <question>.

The <QUESTION> is:
{question}

Furthermore, you already analysed reports and extracted the following passages of relevant information the question is looking for:
---
{examples}
---

Your task is to create a short {length} word explanation for which details the question is asking for.
Make sure to make use of the passages by not directly referencing them but using them to influence the details that might be of help.

Start the answer with 'We search for details on'. Don't mention the question itself in the text.
Your answer:
\end{lstlisting}
\caption{\label{lst:prompt_spec_case_noQ} Prompt for creating the expert-informed information retrieval explanation \textbf{without} the question.}
\end{figure*}

\section{Details on the Experimental Setup}\label{app:experimental_setup}
Following the Information Retrieval Explanation (see Appendix \ref{app:IR_queries}), we also choose to set a relevance threshold for the base setup of our evaluation. For the base evaluation, the threshold is 2 or higher. Again, we argue that for the binary label at hand (relevant or not), the label of relevance 1 might be confusing since in its definition it is not entirely clear whether the source is really relevant. Thus, future investigations should focus on determining uncertainty around relevance labeling.

Furthermore, in the base setup, we use a random baseline, BM25, DRAGON+, GTE-base, ColBERTv2, OpenAI's text-embedding-ada-002, text-embedding3-small, and text-embedding3-large to embed questions, definitions, and paragraphs.

We aggregate the results over all embeddings or top-k values to compare scenarios. Furthermore, we use the text-embedding3-large to showcase single aspects.

\section{Comparing Retrieval with Questions, Definitions and Concepts vs. Explanations}\label{app:quest_def_concept_results}
Table \ref{tab:compare_quest_def_concepts} shows the results of comparing the retrieval with text-embedding-3-large with questions, definitions, and concepts along all metrics and top k values. It becomes apparent that using the sole question for information retrieval is the best.

This might raise the question of whether the definition and concepts are wrong. However, we argue that the definition and concepts work worse for two reasons. First, the definitions and concepts are an aid for the individual labeler to remain consistent with herself. This means the labeler might not explicitly state exact details in the definitions or concepts. The real labeling knowledge may remain with the expert. This is also highly interconnected with the second reason. Neither the definitions nor the concepts were optimized for the search with embeddings. The labeler has a high degree of freedom regarding how long the definitions or concepts are. 

In contrast, the generic and expert-informed explanations are the result of a thought concept to optimize embedding search. As Tables \ref{tab:query_gen_inf3_60} and \ref{tab:query_gen_infall_150} show, these explanations offer dense mentioning of targeted contents relating to the question. They have a higher level of specificity when compared to the example definition and concepts in Table \ref{tab:compare_quest_def_concepts}. 

We argue that this is also the reason why using an example-informed, that is, an expert-informed explanation, improves the retrieval in contrast to using the definitions and concepts of the labelers (see \ref{fig:compare_found}). This is also reinforced by comparing the $generic$ definition with the $informed$ explanations. Interestingly, a small nuance becomes apparent when comparing $inf_{3}$ and $inf_{all}$. There seems to be no significant jump in performance when letting the definition be inspired by three vs. all reports' relevant sources as examples. This indicates that (1) designing the definitions based on a limited sample is enough and (2) there might even be an overfitting in only orientating on examples.

We argue the level of detail of the explanations can serve as a good basis for future definitions and concepts enabling an iterative expert-machine-integrated process. This could ultimately aim to provoke the human to be more precise and reflect with the machine.

\begin{table*}[]
\centering
\begin{tabular}{lllll}
\hline
Setup      & Top K & Found Rel. Sources & Rel. Retrieved Sources & F1-Score        \\ \hline
Question   & 5     & \textbf{0.2263}    & \textbf{0.1503}        & \textbf{0.1806} \\
Definition & 5     & 0.1818             & 0.1208                 & 0.1452          \\
Concepts   & 5     & 0.1960             & 0.1302                 & 0.1565          \\ \hline
Question   & 10    & \textbf{0.3394}    & \textbf{0.1128}        & \textbf{0.1693} \\
Definition & 10    & 0.2909             & 0.0966                 & 0.1451          \\
Concepts   & 10    & 0.3091             & 0.1027                 & 0.1542          \\ \hline
Question   & 15    & \textbf{0.4202}    & \textbf{0.0931}        & \textbf{0.1524} \\
Definition & 15    & 0.3818             & 0.0846                 & 0.1385          \\
Concepts   & 15    & 0.4040             & 0.0895                 & 0.1465          \\ \hline
\end{tabular}
\caption{Recall@K, Precision@K, F1-Score@K for the retrieval with question, definition, and concepts (Embeddings = "text-embedding-3-large").}
\label{tab:compare_quest_def_concepts}
\end{table*}

\section{Aggregated Results for Question vs. No Question}\label{app:Q_v_noQ}
As Table \ref{tab:Q_v_noQ} shows,  the most promising strategy remains expert-informed explanations that exclude the question across all settings. This observation is consistent with the single observation with text-embedding-3-large.

\begin{table*}
\centering
\begin{tabular}{lrrrr}
\hline
question & generic & inf\_3 & inf\_all & expert-informed strategy \\
\hline
0.132 & 0.131 & 0.145 & 0.144 & short\_Q \\
0.132 & 0.130 & 0.142 & 0.142 & long\_Q \\
0.132 & 0.140 & 0.159 & \textbf{0.160} & short\_noQ \\
0.132 & 0.138 & \textbf{0.160} & 0.159 & long\_noQ \\
\hline
\end{tabular}
\caption{Table R.2: Results for the F1-score of the different strategies for optimizing the embedding search across all topk values (5, 10, 15) and embedding models. The best-performing strategies for optimizing the embedding search (in bold) are using expert-informed explanations excluding the question.}
\label{tab:Q_v_noQ}
\end{table*}

\section{All Results with Metrics, Emdeddings and Relevance Thresholds for text-embedding-3-large}\label{app:other_results}
To solidify the results of our experiments, we employ a set of different metrics. In this section, we show the results for the strongest embedding model, text-embedding-3-large. 





\begin{figure}[t]
    \centering	\includegraphics[width=0.48\textwidth]{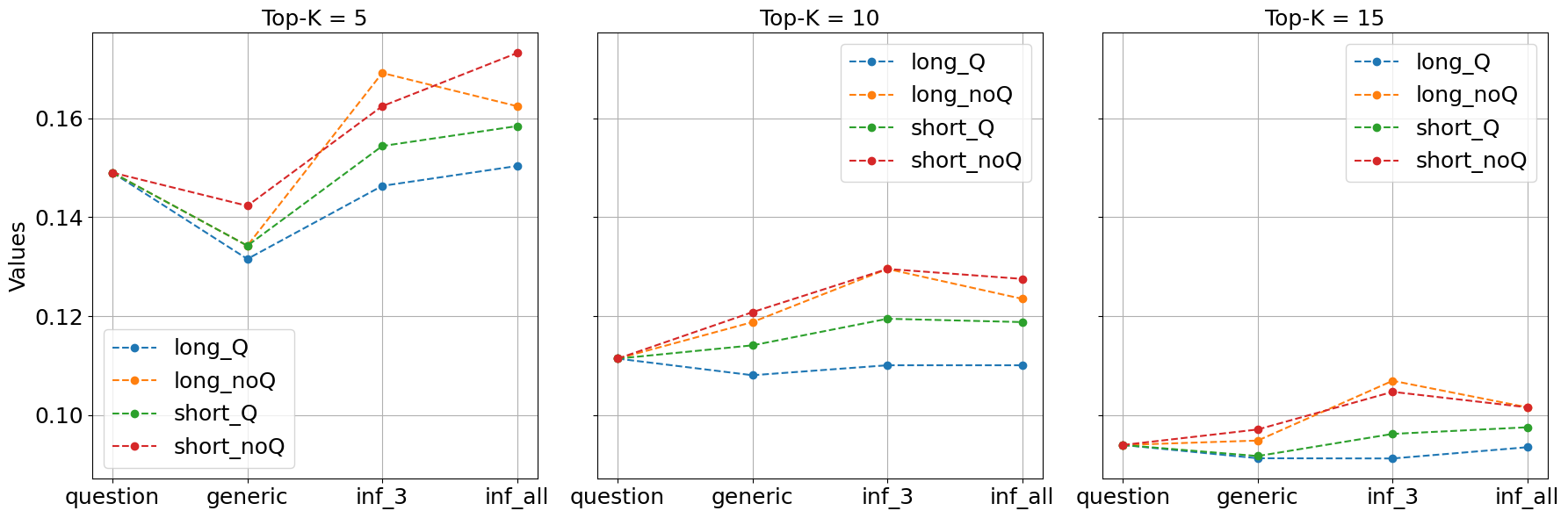}
	\caption{Precision@K for the different experimental setups (Embeddings = "text-embedding-3-large").}
	\label{fig:ratio_retrieval_2}
\end{figure}

\begin{figure}[t]
    \centering	\includegraphics[width=0.48\textwidth]{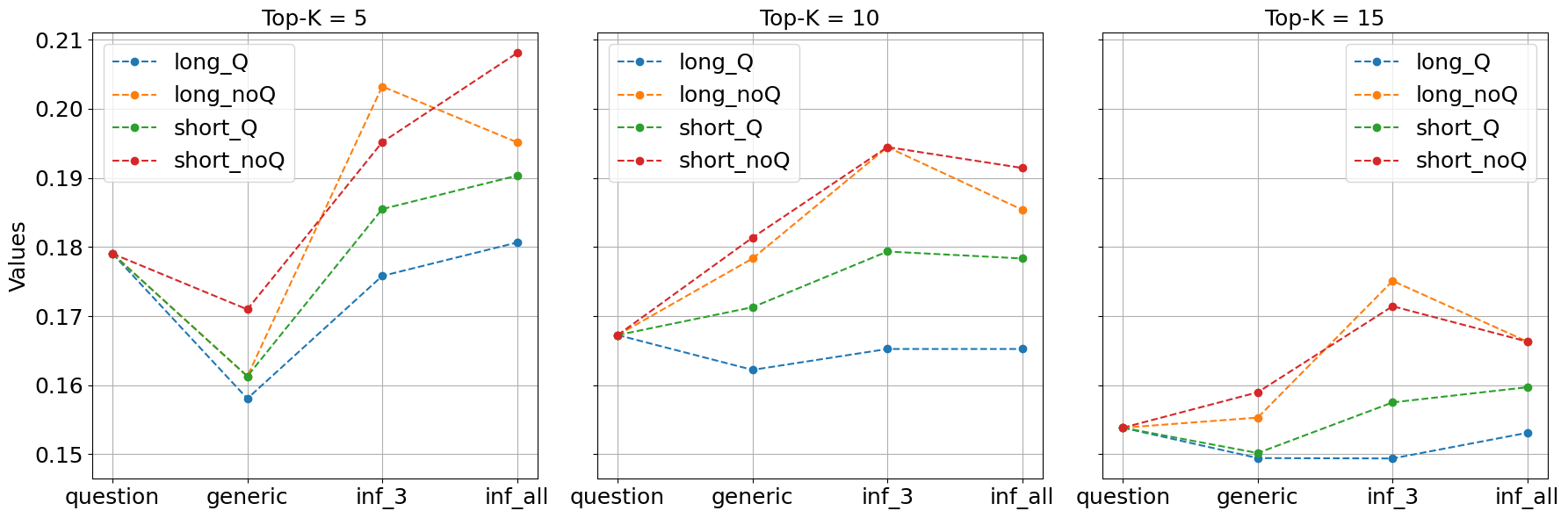}
	\caption{F1-score for the different experimental setups (Embeddings = "text-embedding-3-large").}
	\label{fig:f1_base}
\end{figure}

While the results in Figures \ref{fig:ratio_retrieval_2} and \ref{fig:f1_base} confirm the results in Figure \ref{fig:compare_found}, they add one dimension of nuance. The results indicate that a higher top k value is optimal because more annotated sources are found. However, it also comes with the downside of more irrelevant sources as well. These results again indicate that more nuanced relevant labels abstracting from fixed thresholds might be optimal. 

\begin{figure}[t]
    \centering	\includegraphics[width=0.48\textwidth]{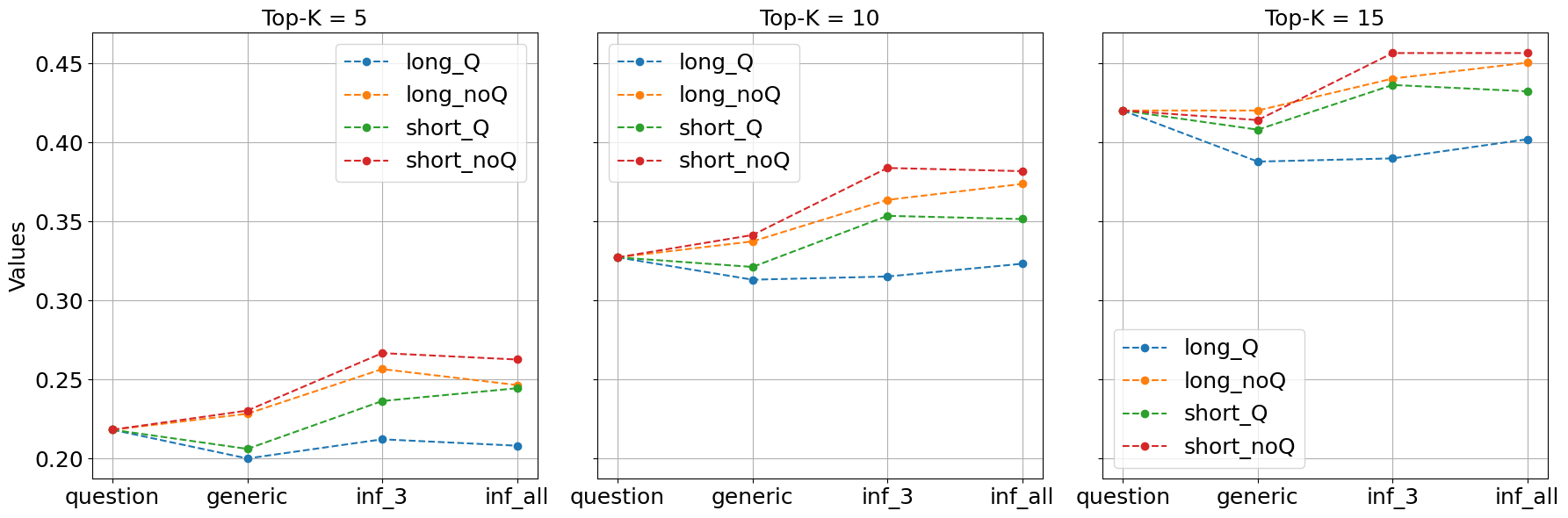}
	\caption{Recall@K for the different experimental setups (Embeddings = "text-embedding-3-small").}
	\label{fig:recall_small}
\end{figure}

\begin{figure}[t]
    \centering	\includegraphics[width=0.48\textwidth]{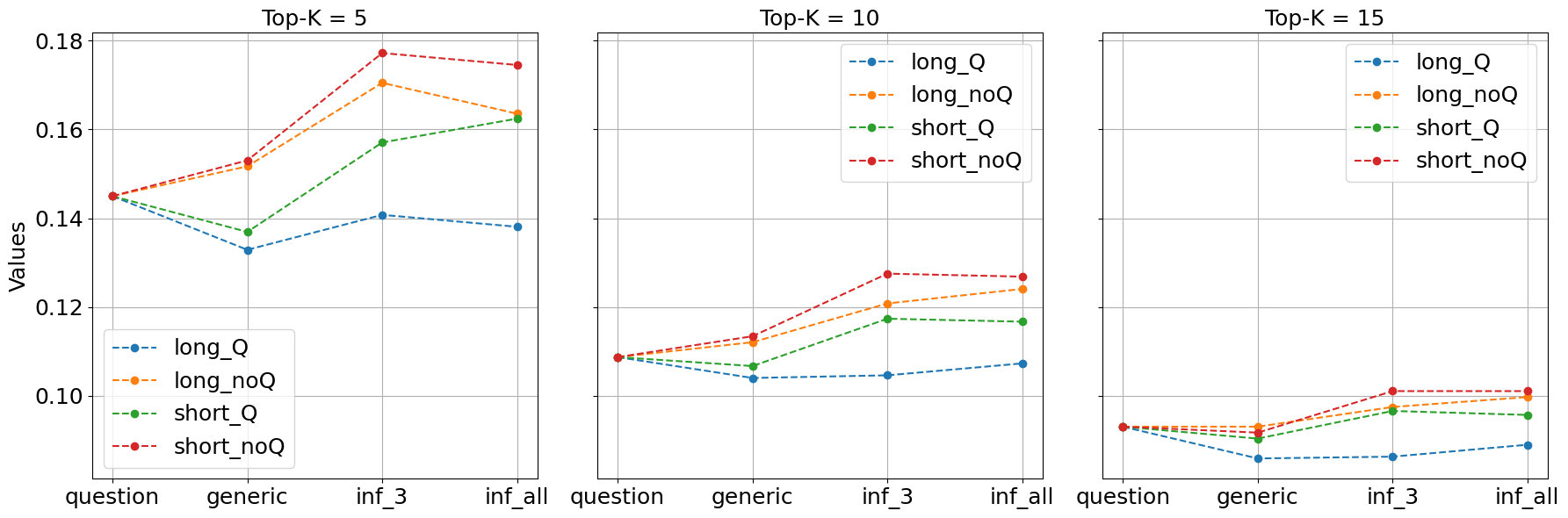}
	\caption{Precision@K for the different experimental setups (Embeddings = "text-embedding-3-small").}
 \label{fig:precision_small}
\end{figure}

\begin{figure}[t]
    \centering	\includegraphics[width=0.48\textwidth]{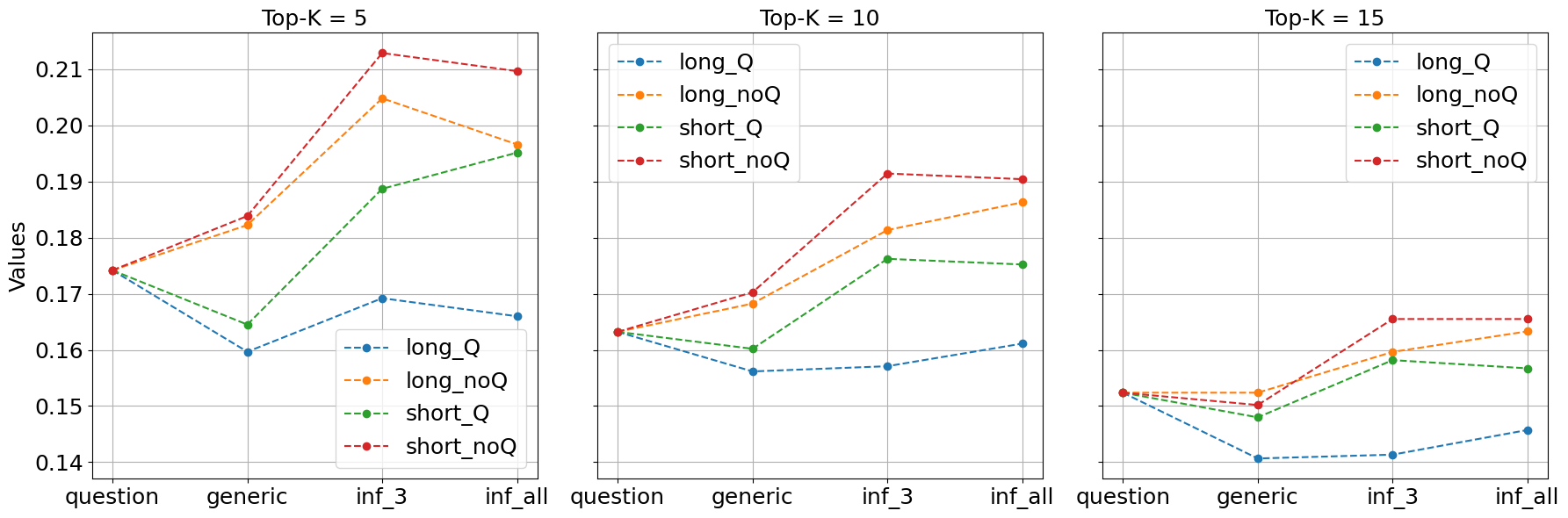}
	\caption{F1-score for the different experimental setups (Embeddings = "text-embedding-3-small").}
 \label{fig:f1_small}
\end{figure}

Furthermore, it is interesting to see how the results change when changing the underlying embedding model. Thus, we also change the embedding model from "text-embedding-3-large" to "text-embedding-3-small". Again, the results stay vastly the same (see Figures \ref{fig:precision_small}, \ref{fig:recall_small}, and \ref{fig:f1_small}). However, "text-embedding-3-small" scores are consistently a bit lower. This is in line with their general capabilities.\footnote{A comparison can be found here: \url{https://platform.openai.com/docs/guides/embeddings/embedding-models}.}

\begin{figure}[t]
    \centering	\includegraphics[width=0.48\textwidth]{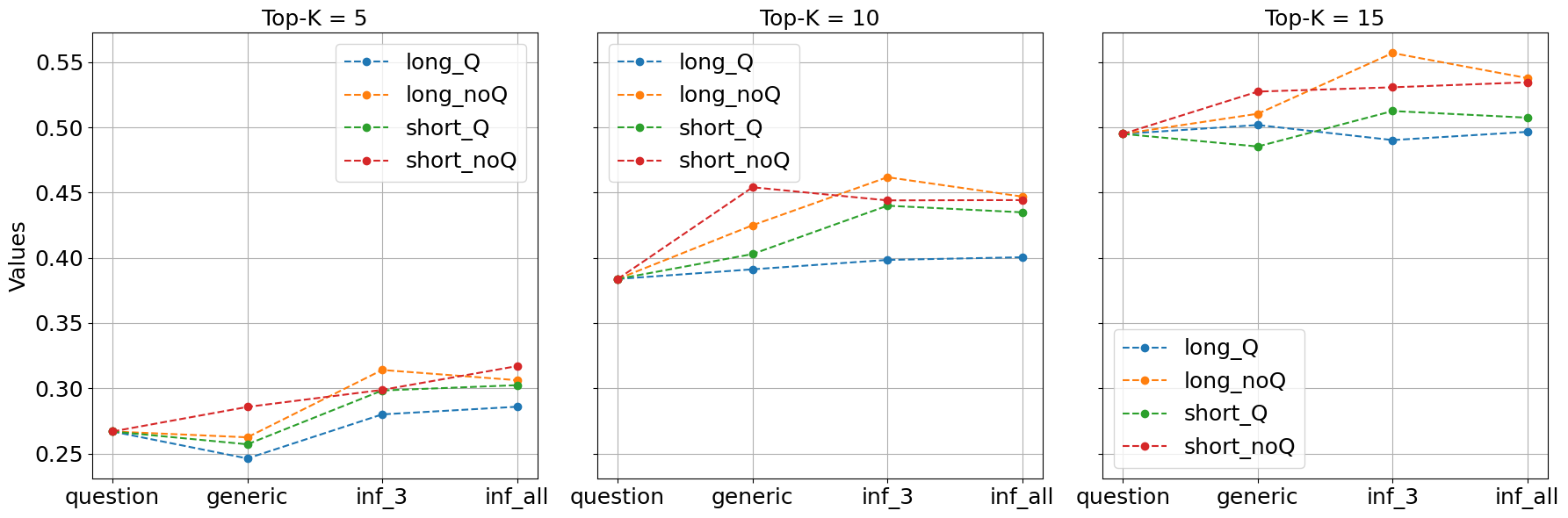}
	\caption{Recall@K for the different experimental setups and a relevance threshold of 1 (Embeddings = "text-embedding-3-large").}
	\label{fig:compare_found_1}
\end{figure}

\begin{figure}[t]
    \centering	\includegraphics[width=0.48\textwidth]{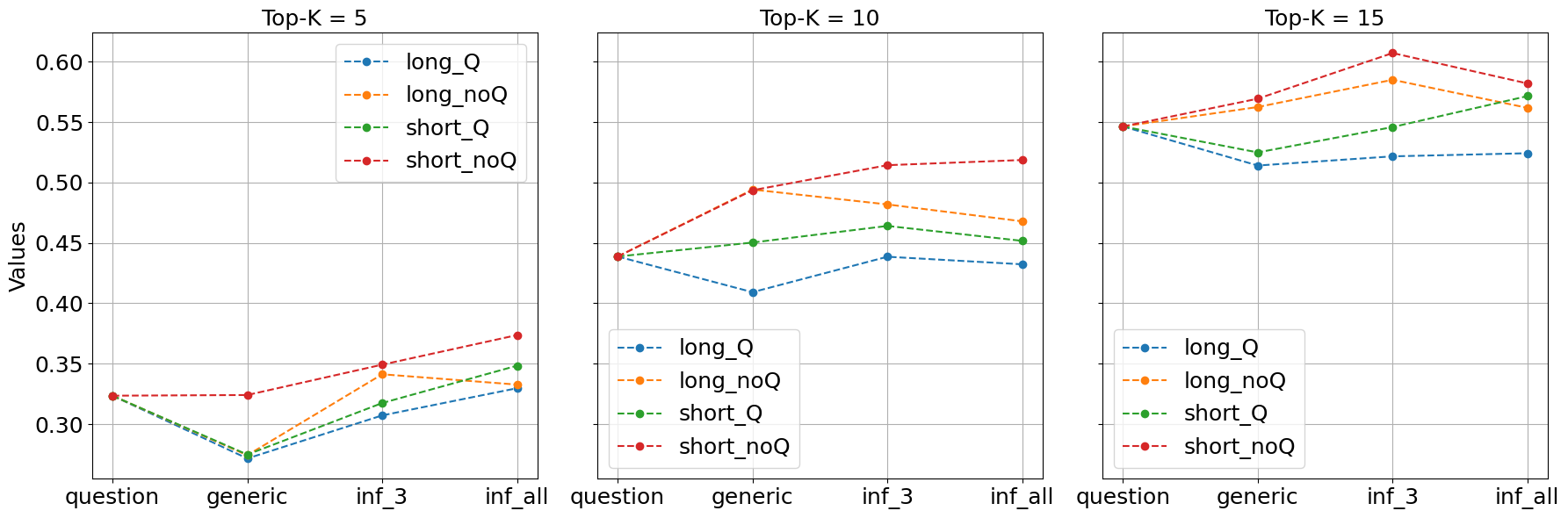}
	\caption{Recall@K sources for the different experimental setups and a relevance threshold of 3 (Embeddings = "text-embedding-3-large").}
	\label{fig:compare_found_3}
\end{figure}

Finally, we choose the relevance threshold to be 2 for all our experiments. Again, the results are consistent when changing the threshold to 1 or 3 (see Figures \ref{fig:compare_found_1} and \ref{fig:compare_found_3}). Collectively, these results suggest that the findings are solid.

\begin{figure}[t]
    \centering	\includegraphics[width=0.48\textwidth]{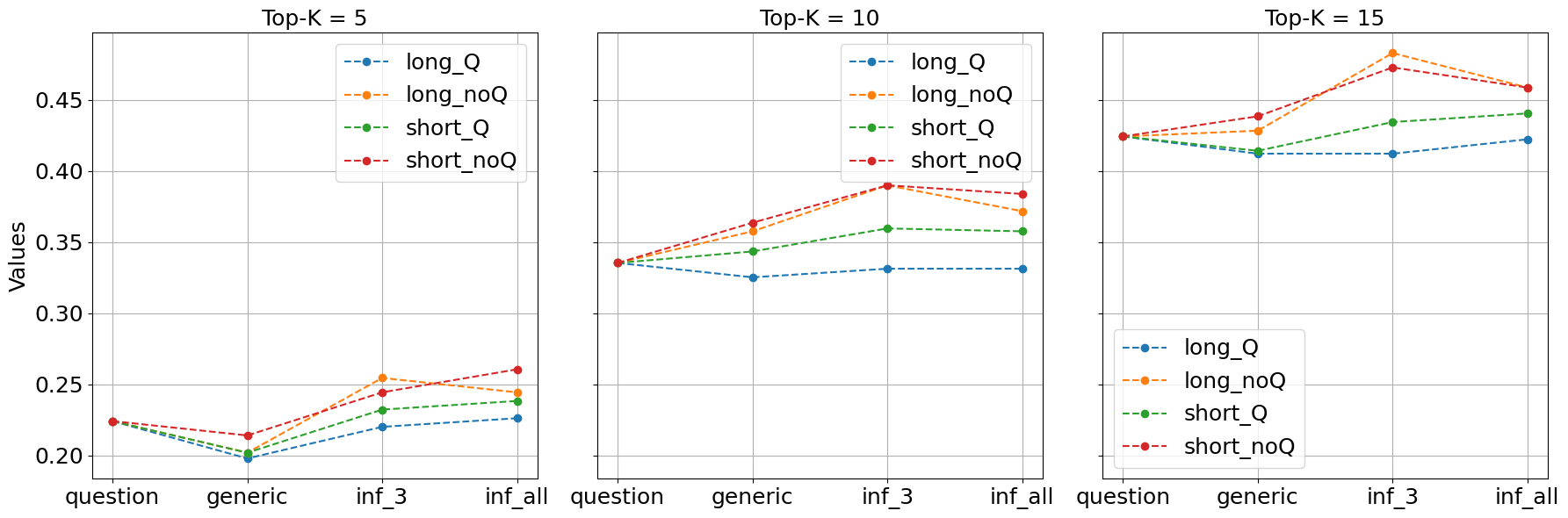}
	\caption{Recall@K for the different experimental setups (Embeddings = "text-embedding-3-large").}
	\label{fig:recall_base}
\end{figure}

\end{document}